\documentclass[aps,pra,twocolumn,showpacs,epsfig,superscriptaddress]{revtex4-1}
\usepackage[latin1]{inputenc}
\usepackage{color,graphicx}
\usepackage{amsmath,amssymb,amsthm,dsfont}

\usepackage{setspace}
\usepackage{epsfig}
\usepackage{epstopdf}
\usepackage{color}
\newcommand{\unit}{1\!\!1}

\newcommand{\bra}[1]{\left\langle#1\right\vert}
\newcommand{\ket}[1]{\left\vert#1\right\rangle}

\begin{document}

\title{Quantum-state transfer via resonant tunnelling through local field induced barriers}

\author{S.~Lorenzo}
\affiliation{Dip.  Fisica, Universit\`a della Calabria, 87036 Arcavacata di Rende (CS), Italy}
\affiliation{INFN - Gruppo collegato di Cosenza}

\author{T.~J.~G.~Apollaro}
\affiliation{Dip.  Fisica, Universit\`a della Calabria, 87036 Arcavacata di Rende (CS), Italy}
\affiliation{INFN - Gruppo collegato di Cosenza}
\affiliation{Centre for Theoretical Atomic, Molecular and Optical Physics,
School of Mathematics and Physics, Queen's University, Belfast BT7 1NN, United Kingdom}

\author{A.~Sindona}
\affiliation{Dip.  Fisica, Universit\`a della Calabria, 87036 Arcavacata di Rende (CS), Italy}
\affiliation{INFN - Gruppo collegato di Cosenza}

\author{F.~Plastina}
\affiliation{Dip.  Fisica, Universit\`a della Calabria, 87036 Arcavacata di Rende (CS), Italy}
\affiliation{INFN - Gruppo collegato di Cosenza}
\pagenumbering{arabic}

\date{\today}

\begin{abstract}
Efficient quantum-state transfer is achieved in a uniformly
coupled spin-$1/2$ chain, with open boundaries, by application of
local magnetic fields on the second and last-but-one spins,
respectively. These effective \textit{barriers} induce appearance
of two eigenstates, bi-localized at the edges of the chain, which
allow a high quality transfer also at relatively long distances.
The same mechanism may be used to send an entire e-bit~(e.g., an
entangled qubit pair) from one to the other end of the chain.
\end{abstract}

\maketitle

\section{Introduction}
Quantum State Transfer~(QST), i.e., the reliable transfer of an
arbitrary quantum state between different quantum processing
units, is one of the major tools of distributed quantum computing
and provides the basic `building block' for any quantum
communication protocol~\cite{Nielsen,interkimble}. When the
information is encoded in intrinsically localized units, an
efficient quantum communication channel can be realized with
effective spin systems~\cite{Bose2003}, in order to avoid the
difficult problem of interfacing with flying qubits. This channel
becomes especially useful for short ranged, on-chip
communication~(see Ref.~\cite{channels} and references therein).

For the QST of one qubit~(which may be part of an entangled or,
more generally, a correlated pair~\cite{campbell}), a number of
protocols have been described employing spin-$\frac{1}{2}$ chains
as quantum data bus to transfer information between their first
and last spins~(the sender and receiver, respectively). In
particular, a high Fidelity transmission can be obtained if
additional resources are employed with respect to the original
plain scheme of Ref.~\cite{Bose2003}. Examples include the
encoding of quantum states on spatially extended wave
packets~\cite{PaganelliGP09,LindenO04}, the use of local end-chain
operations~\cite{DiFrancoPK}, of local memories and parallel
quantum channels~\cite{GiovannettiBB}, or of protocols employing
time-dependent interactions~\cite{timedependent}. A perfect state
transfer, which is unattainable in a uniformly coupled chain, can
be achieved instead by a proper pre-engineering of the coupling
strengths. The key advantage in this case is that no external time
dependent controls are needed, as the transfer is realized through
the intrinsic dynamics of the chain. Perfect QST, which may be
thought of as a particular instance of a more generic swap
operation~\cite{swap}, is entailed by accurate settings of the
intra-channel coupling strengths giving rise to a linear
dispersion relation for excitations propagating across the
channel~\cite{Christandl2004}. However, dispersion during
transmission occurs in most spin chains due to the nontrivial
structure of the many-body Hamiltonian describing the channel, and
the design of a non-dispersive channel requires a demanding
engineering of the Hamiltonian parameters. A systematic analysis
on how to set the couplings to allow for a perfect state transfer
can be found in Refs.~\cite{vinet85,bruderer}.

On the other hand, a {\it quasi perfect} transfer~\cite{vinet} can
be obtained by modifying only a few couplings of an otherwise
homogeneous quantum channel~\cite{CamposVenutiET08}, in order to
obtain a ballistic excitation transfer~\cite{BACVV}, or Rabi-like
oscillations between eigenstates having support only on the sender
and receiver
sites~\cite{Wojcik2005,Gualdi08,Plastina2007,PaganelliPG2006,heule,linne}.

In this work, we propose a new transfer protocol of the latter
kind and analyze the efficiency and reliability of state
transmission in presence of a minimal engineering, which depends
on the resonant tunnelling of spin excitations induced by
application of local magnetic fields near the sending and
receiving sites. Specifically, we require the sender and receiver
to have access and control over the local fields applied on their
neighboring spins, which are increased by $\omega$ with respect to
the rest of the chain~(see the sketch in Fig.~\ref{chainw}). As
discussed in Ref.~\cite{Plastina2007,PaganelliPG2006,apo06}, in an
open spin-$1/2$ chain of $N$ nodes, these extra local fields
induce appearance of two single-particle states, which are
`bi-localized' on sites $2$ and $N-1$ and can be exploited to
perform QST between them~\cite{heule,linne} in a time
$t{\sim}\omega^{N-2}$. However, this is not the only effect
produced by the local fields. The geometric confinement, due to
the open boundary conditions imposed on the chain, induce
appearance of a further pair of eigenstates which are localized on
the first and last sites and can be exploited for a much faster
QST. Indeed, once the spin chain is fermionized via the
Jordan-Wigner transformation, it is easy to recognize that the
local fields create effective potential barriers for the
single-particle excitations. If these barriers have equal
heights~(thus establishing a mirror symmetry~\cite{shi}), a
coherent resonant tunnelling occurs between the first and last
sites, giving rise to information transfer.

The paper is organized as follows: in Sec~\ref{Sec.Model} the
model with the magnetic field `barriers' is solved, and the
appearance of the bi-localized states mentioned above is
discussed; in Sec.~\ref{Sec.Conf} the transmission Fidelity is
studied and the effectiveness of the local fields allowing for a
very high quality QST is demonstrated. Furthermore, in
subsection~\ref{SubSec.Noise} the resilience with respect to noise
is analyzed, while  in subsection~\ref{SubSec.Twoqubit} the
possibility of transferring more than one qubit is briefly touched
upon. After that, in Sec.~\ref{Sec.Time}, a time-dependent
protocol based on the switching of the local fields is presented
and, finally, some concluding remarks are drawn in
Sec.~\ref{Sec.Conc}.

\section{The model and its properties}
\label{Sec.Model} We consider a linear chain of spin-${1}/{2}$
particles residing at sites, $n{=}1,2,..., N$, in a lattice of
unit lattice constant. The $N$ spins are coupled through the
homogeneous nearest-neighbor $XX$ model
\begin{equation}\label{H-general-local}
\mathcal{H}=-J \left \{ \dfrac{1}{2} \sum_{n{=}1}^{N{-}1}
(\sigma_{n}^x\sigma_{n{+}1}^x{+}\sigma_{n}^y\sigma_{n{+}1}^y) +
\sum_{n=1}^N K_n \sigma_{n}^z \right \},
\end{equation}
here expressed in $\hbar{=}1$-units, which will be used throughout
this paper. In Eq.~\eqref{H-general-local},
$\sigma_{n}^{\alpha}$~($\alpha{=}x,y,z$) are the usual Pauli
matrices for the spin at the $n$-th site, probed by a local
magnetic field of intensity $K_n$, and $J$ is the exchange
coupling strength between two nearest neighboring sites. In the
following $J$ will be set to $1$ and taken as our energy unit
(therefore, times will be given in $1/J$ units).

As in the protocol of Ref.~\cite{Bose2003}, we begin with the
chain being prepared with all spins up, say,  in  the  initial
state $\ket{\textbf{0}}{=}\ket{0}^{\otimes N}$ in which $\ket{0}$
and $\ket{1}$ denote the spin-up and down states along the $z$
axis, respectively. Next, we initialize the first spin of the
chain to the state $\ket{\psi_{in}}{=}\alpha\ket{0}+\beta \ket{1}$
and let the chain follow the time-evolution generated by the
Hamiltonian~(\ref{H-general-local}). Since
$[\mathcal{H},\sum_{n=1}^{N}\sigma_z^n]{=}0$, the dynamics take
place in the invariant subspaces with $0$ and $1$ flipped spins,
where the former is made up of the state $\ket{\textbf{0}}$ alone,
while the latter is spanned by the computational basis states
$|\textbf{j}\rangle=\sigma_{j}^{+}|\textbf{0}\rangle \equiv
|0_1,0_2, \ldots, 0_{j-1},1_j,0_{j+1}, \dots \rangle$.

The state of the last spin, $\rho_N(t)$, is obtained from the time
evolved state of the chain by tracing out all but the $N$-th spin,
and the aim of the QST protocol is to retrieve the state encoded
in the first spin from the last one. The efficiency of the state
transfer is then quantified by the Fidelity
$F(t){=}\bra{\psi_{in}}\rho_N(t)\ket{\psi_{in}}$, which equals $1$
in the case of a perfect transfer. In order to evaluate the
channel quality independently of the specific input state, we
refer to the average Fidelity ${\overline{F}}(t)$ by integrating
$F(t)$ over all possible pure input states of a qubit. This leads
to
\begin{equation}\label{FidMean}
{\overline{F}}(t)=\frac{|f_{N1}(t)|}{3}+\frac{|f_{N1}(t)|^2}{6} +
\frac{1}{2},
\end{equation}
where $f_{N1}(t)=\langle
\textbf{N}|e^{-i{\cal{H}}t}|\textbf{1}\rangle$ is the transition
amplitude of a spin excitation from the first to the last site of
the chain. In the following, with the term Fidelity we will refer
to the quantity given by Eq.~(\ref{FidMean}).

The same effective channel can be used to transfer entanglement,
with the first spin sharing an initial singlet state with an
external and uncoupled qubit. The amount of (transferred)
entanglement between the last spin of the chain and the external
one at a subsequent time $t$, as measured by the Concurrence, is
given by~\cite{Bose2003}
\begin{equation}\label{ConC}C(t)= |f_{N1}(t)|.\end{equation}
Therefore, in order to perform efficiently both of the tasks,
namely the state and entanglement transfers, it is necessary to
achieve a value of $|f_{N1}(t)|$ as close as possible to $1$ at a
certain time $t^*$.

Because of the time invariance of the subspaces with a given
number of flipped spins, the calculation of $f_{N1}(t)$ is reduced
to diagonalizing the Hamiltonian in the single excitation sector,
where Eq.~(\ref{H-general-local}) can be expressed as a
tri-diagonal matrix whose elements are $\mathcal{H}^{(1)}_{nm}{=}2
K_n\delta_{nm}{-}\left(\delta_{n,n+1}+\delta_{n,n-1}\right)$.
Indeed, the transition amplitude $f_{N1}$ can be written as
\begin{equation}\label{f1N}
f_{N1}(t){=}\sum_{k=1}^N \langle\mathbf{N}|\mathbf{a}_k\rangle
\langle\mathbf{a}_k|\mathbf{1}\rangle e^{-i \lambda_k t}
\end{equation}
where $\lambda_k$ are the eigenvalues and
$|\mathbf{a}_k\rangle{=}\sum_{j=1}^N a_{kj}|\mathbf{j}\rangle$ the
corresponding eigenvectors of $\mathcal{H}^{(1)}$, arranged in
increasing order, i.e., $\lambda_{k^{\prime}}{>} \lambda_k$ for
$k^{\prime}{>}k$.

As we will show below, a large value for $|f_{N1}|$  can be
obtained by modifying only two local fields in such a way that
only two eigenvectors among the $|\mathbf{a}_k\rangle$'s have a
non-negligible superposition with $\ket{\mathbf{1}}$ and
$\ket{\mathbf{N}}$. Correspondingly, the time evolution induced by
${\cal H}$ gives rise to an effective Rabi oscillation of the
spin-excitation between the first and the last sites of the chain.

\begin{figure}[ht!]
    \centering
 \includegraphics[width=\linewidth]{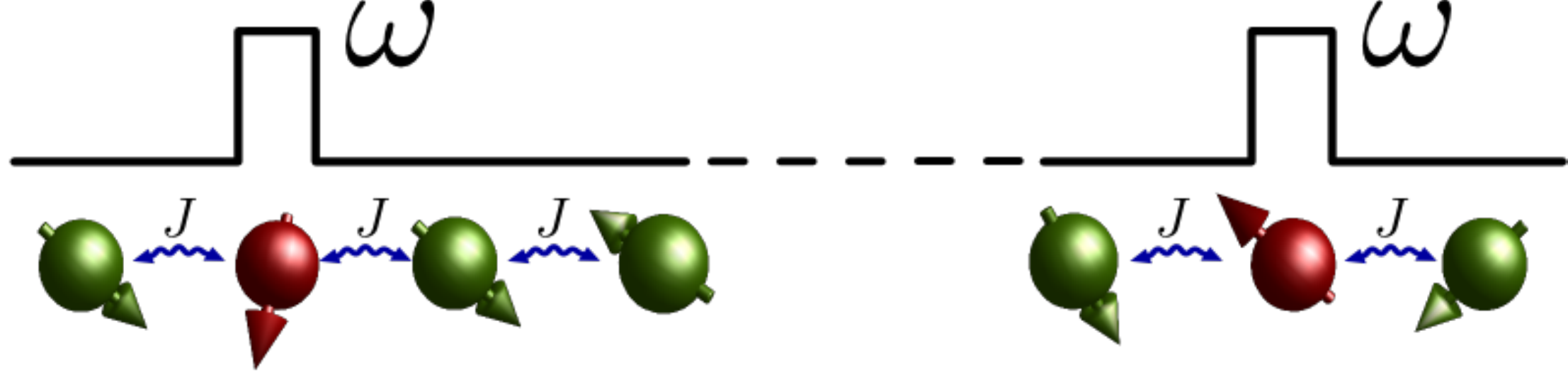}
\caption{(color on-line): Sketch of the spin chain with sender and
receiver located at the first and last sites, and with local field
barriers of height  $\omega$ applied to the second and
last-but-one sites.}
    \label{chainw}
\end{figure}

Specifically, we assume that the local magnetic fields are applied
to the second and last-but-one spins, which in the following will
be denoted as \textit{barrier qubits}, by setting
$K_n{=}\omega\left(\delta_{n,2}+\delta_{n,N-1} \right)$ in
Eq.~(\ref{H-general-local}), which gives rise to the model
depicted in Fig.~\ref{chainw}. This yields an effective decoupling
of the first and the last spins of the chain whose dynamics take
place mainly in a subspace spanned by two particular eigenstates
of $\mathcal{H}^{(1)}$, which are close enough in energy and
bi-localized at the edges of the chain.

\begin{figure}[ht!]
\center{\includegraphics[width=7cm]{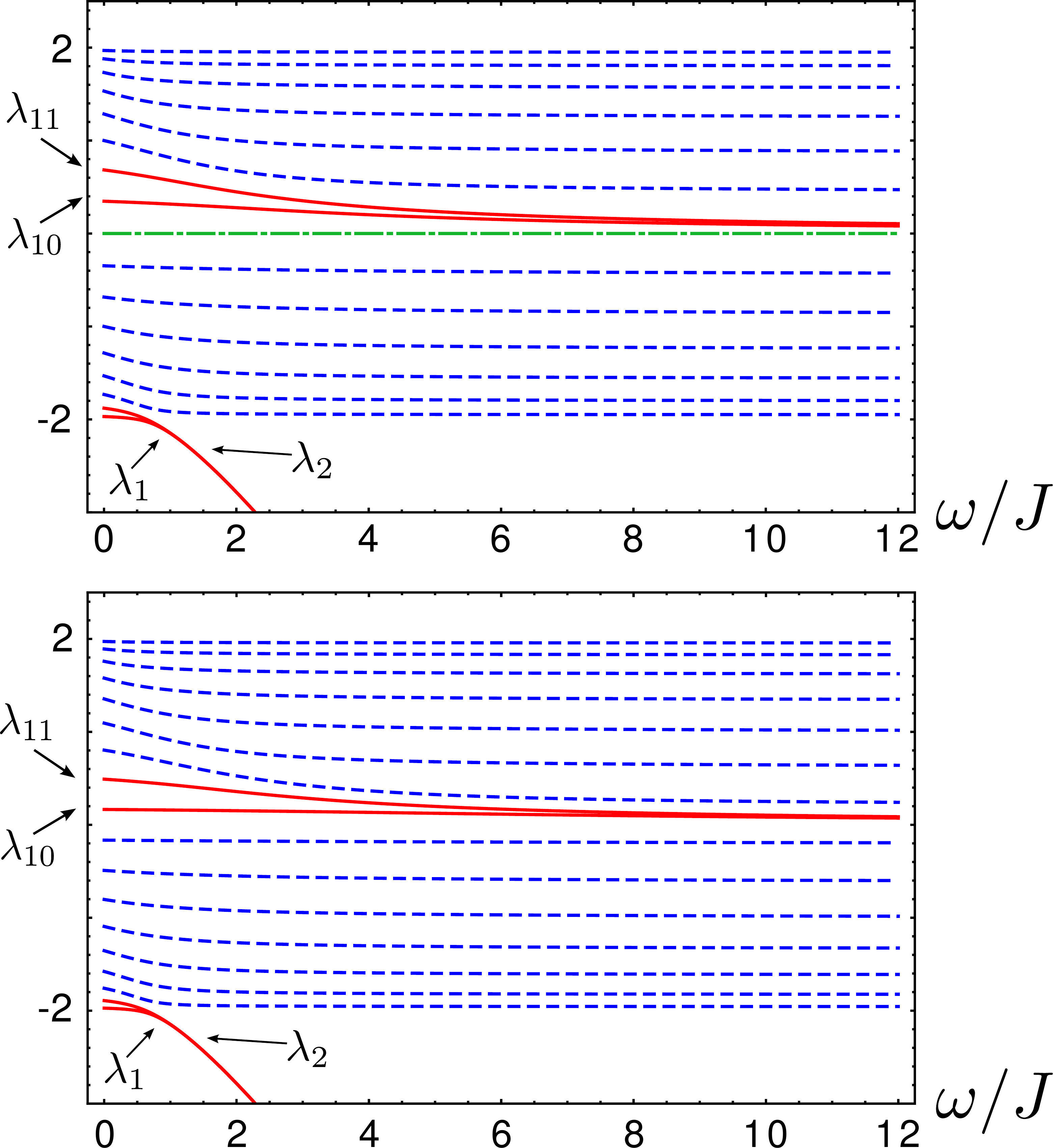}}
\caption{(color online): Spectrum of a spin chain with
$N{=}17$~(upper panel) and $N{=}18$ sites~(lower panel) versus
$\omega$: in both cases, the two lowest eigenenergies move outside
the band as $\omega$ increases, while the two positive
eigenenergies closest to zero become quasi-degenerate. The latter
are represented by red, solid lines. A zero-energy eigenstate
occurs in the odd chain, whose eigenvalue is represented by the
green dot-dashed line. All energies values are reported in units
of $J$.} \label{eige}
\end{figure}

To confirm these expectations, we study the spectrum of
${\cal{H}}^{(1)}$, reported in Fig.~\ref{eige} for spin chains of
$N{=}17,18$ sites. The cases of even and odd site numbers are
analyzed separately, as they display slightly different features.

In order to quantify the localization of the eigenvectors
$\ket{\mathbf{a}_k}$, induced by the magnetic field $\omega$, we
use the Inverse Participation Ratio (IPR), whose application to
state transfer has been discussed in Ref.~\cite{ZwickO2011}, which
is defined as
\begin{equation*}
\mbox{IPR}(\ket{\mathbf{a}_k}){=}\dfrac{\sum_{i=1}^N| a_{ki}|^2}
{\sum_{i=1}^N |a_{ki}|^4}. \label{IPR}
\end{equation*}
When a state is localized on a single site $n$, i.e.,
$a_{ki}{=}\delta_{ni}$, the IPR takes its minimum possible value
$\mbox{IPR}=1$. On the other hand, an extended state distributed
over a large number of sites yields an IPR value of the order of
the chain length. Notice that the IPR gives information about the
degree of localization of a given eigenstate only, but it does not
say anything about its spatial distribution~(with the exception of
the $\mbox{IPR}{=}N$ case, corresponding to a state uniformly
spread over the whole system).

\begin{figure}[ht!]
\center{\includegraphics[width=7cm]{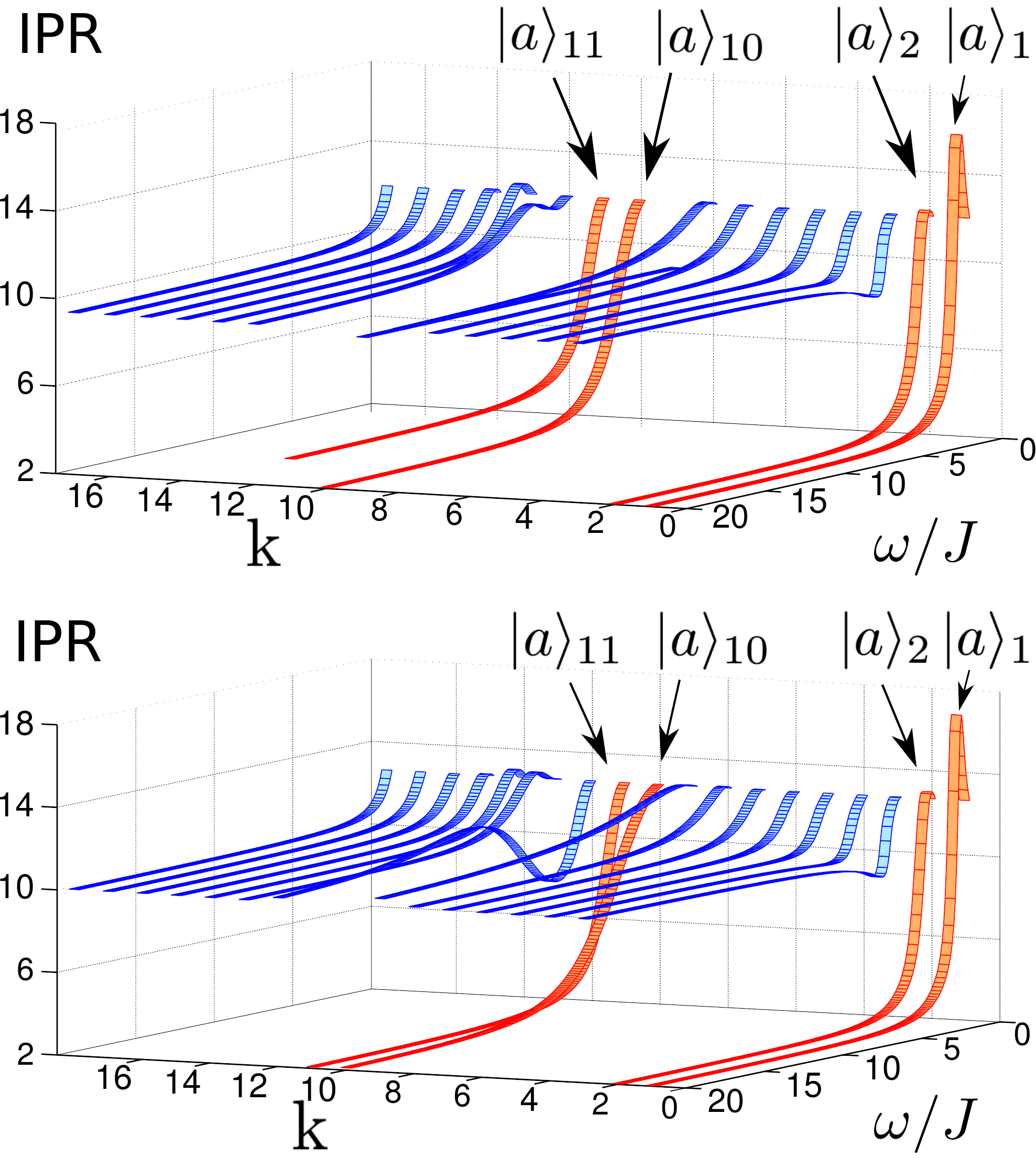}} \caption{(color
online): $\mathsf{IPR}$ for the eigenstates $\ket{a}_k$ of a chain
of $N{=}17$ (upper plot) and $N{=}18$ (lower plot) sites, sorted
by increasing eigenvalues, versus $\omega$ (in units of $J$). The
first two eigenstates, $\ket{a}_1$ and $\ket{a}_2$, rapidly reach
$IPR{\simeq}2$, becoming bi-localized on the barrier qubits . The
states corresponding to $k=10,11$, which bi-localize on the sender
and receiver qubits, reach the value $\mbox{IPR}{\simeq}2$ for the
even chain, whereas one of them remains slightly above that value
for the odd chain.} \label{eigV}
\end{figure}

In Fig.~\ref{eigV}, we report the IPR of the eigenstates, ordered
by ascending eigenvalues, for $N=17$ and $18$. The effect of
increasing $\omega$ is twofold. First, it causes a strong
localization of the two eigenvectors $\ket{\mathbf{a}_{1,2}}$:
IPR$(\ket{\mathbf{a}_{1,2}}){\simeq}2$. These are the two
lowest-lying eigenvalues, emerging out the unperturbed
($\omega{=}0$)~energy band~$\lambda_k{\in}(-2,2)$ (see
Fig.~\ref{eige}). By increasing $\omega$, these states localize on
the two barrier qubits and therefore their contribution to the
quantity in Eq.~(\ref{f1N}) is negligible. Second, another pair of
eigenvectors is found, with positive energies close to zero, which
reduce their IPR to a value asymptotically tending to $2$ for even
site numbers~(Fig.~\ref{eigV}{\bf b}), while remaining slightly
above $2$ for odd site numbers~(Fig.~\ref{eigV}{\bf a}).

The localization properties of these eigenstates are crucial for
quantum-information transfer as they turn out to give the main
contributions in Eq.~(\ref{f1N}). The remaining intra-band
eigenstates hold their extended nature and, for even $N$, they
have a negligible superposition with the states
$\left\{\ket{\mathbf{1}},\ket{\mathbf{N}}\right\}$, so that the
dynamics occur in an effective two-level subspace. On the other
hand, in the odd-$N$ case, an eigenvector with zero energy
eigenvalue is present, which, independently of $\omega$, has a
constant amplitude on the sender and receiver sites, given by
$\sqrt{\frac{2}{N+1}}$ . As a consequence, its contribution to
Eq.~(\ref{f1N}) cannot be neglected for short chains, and the
resulting  effective dynamics involve three levels.
Furthermore, from Fig.~\ref{eige} we see that other intra-band
eigenvalues experience a downward shift and the eigenvalues of the
bi-localized states become quasi-degenerate with energies close to
zero.

\noindent  With these results at hand, we are now in position to
evaluate the transition amplitude $f_{N1}$, and then the
Fidelity~(\ref{FidMean}) and the Concurrence~\eqref{ConC}.

\section{Figures of merit for the transmission}
\label{Sec.Conf} 
\begin{figure}[ht!]
\center{{\bf (a)}\hskip4cm{\bf (b)}}
\center{
       \includegraphics[width=\linewidth] {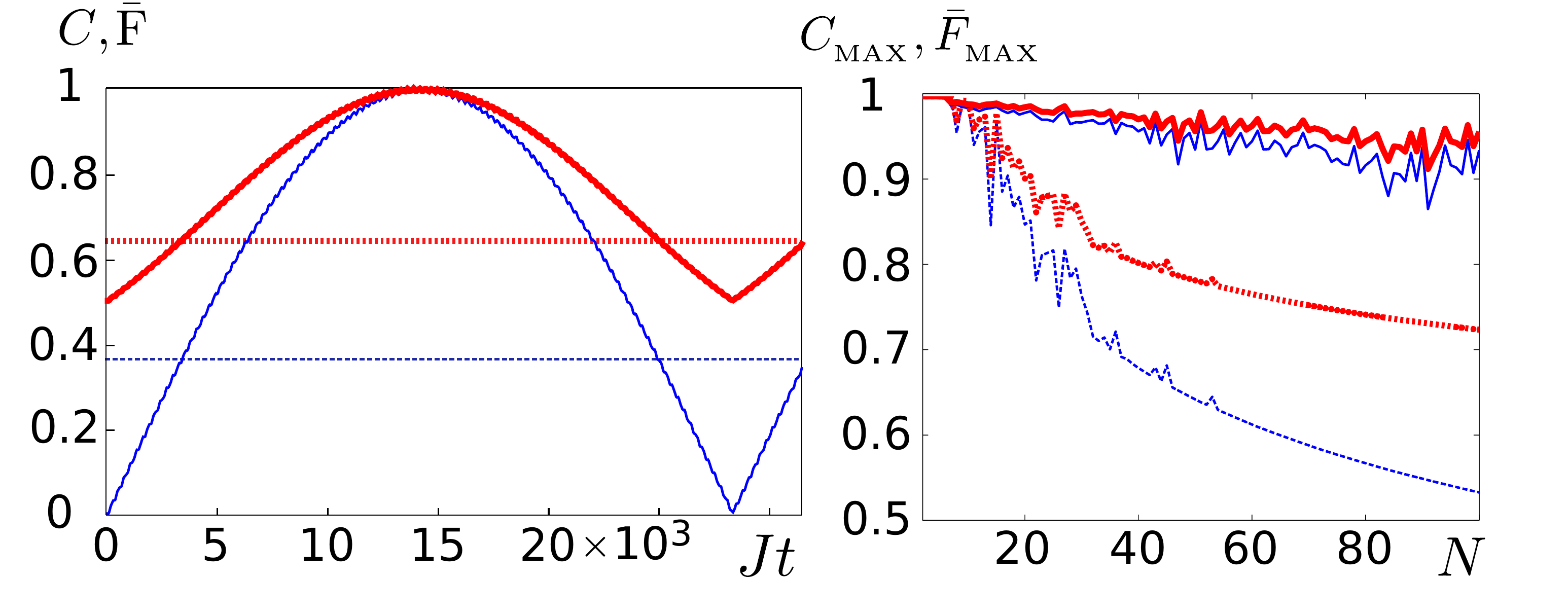}}
\caption{(color online): {\bf a}) Fidelity (red thick line) and
Concurrence (blue thin line, starting from $C=0$ at $t=0$) for a
chain of $N=100$ spins with $\omega{=}100 J$. The dashed red and
blue lines are, respectively, the maximum value of the Fidelity
and of the Concurrence attainable for the homogeneous chain with
$\omega{=}0$. Time is expressed in units of $J^{-1}$. {\bf b})
Maximum of Fidelity (red thick line) and Concurrence (blue thin line) versus the number
of sites $N$ for $\omega{=}10 J$ (solid line) and $\omega{=}0$
(dashed line)} \label{CF100}
\end{figure}

The average transmission Fidelity and the
transmitted Concurrence are reported in Fig.~\ref{CF100}, both as
functions of time and chain length, for fixed values of the
auxiliary local fields $\omega$ applied to the second and
last-but-one sites. To better appreciate the results, they are
compared with the homogeneous case $\omega{=}0$. In
Fig.~\ref{CF100}~{\bf a} we observe a significant improvement of
Fidelity and Concurrence in  presence of $\omega$ with respect to
the homogeneous case, while Fig.~\ref{CF100}~{\bf b} shows that
the difference becomes more and more pronounced with increasing
the chain length. Indeed, at $\omega{=}0$, many terms enter the
sum~(\ref{f1N}), giving rise to a destructive interference that
rapidly suppresses the transfer efficiency (as measured both by
Fidelity and Concurrence). On the other hand, in presence of the
auxiliary fields $\omega$, only two eigenvectors enter
significantly the transition amplitude $f_{N1}(t)$ so that both
the state and entanglement transfers are of high quality.

\begin{figure}[ht!]
\center{\includegraphics[width=0.8\linewidth]{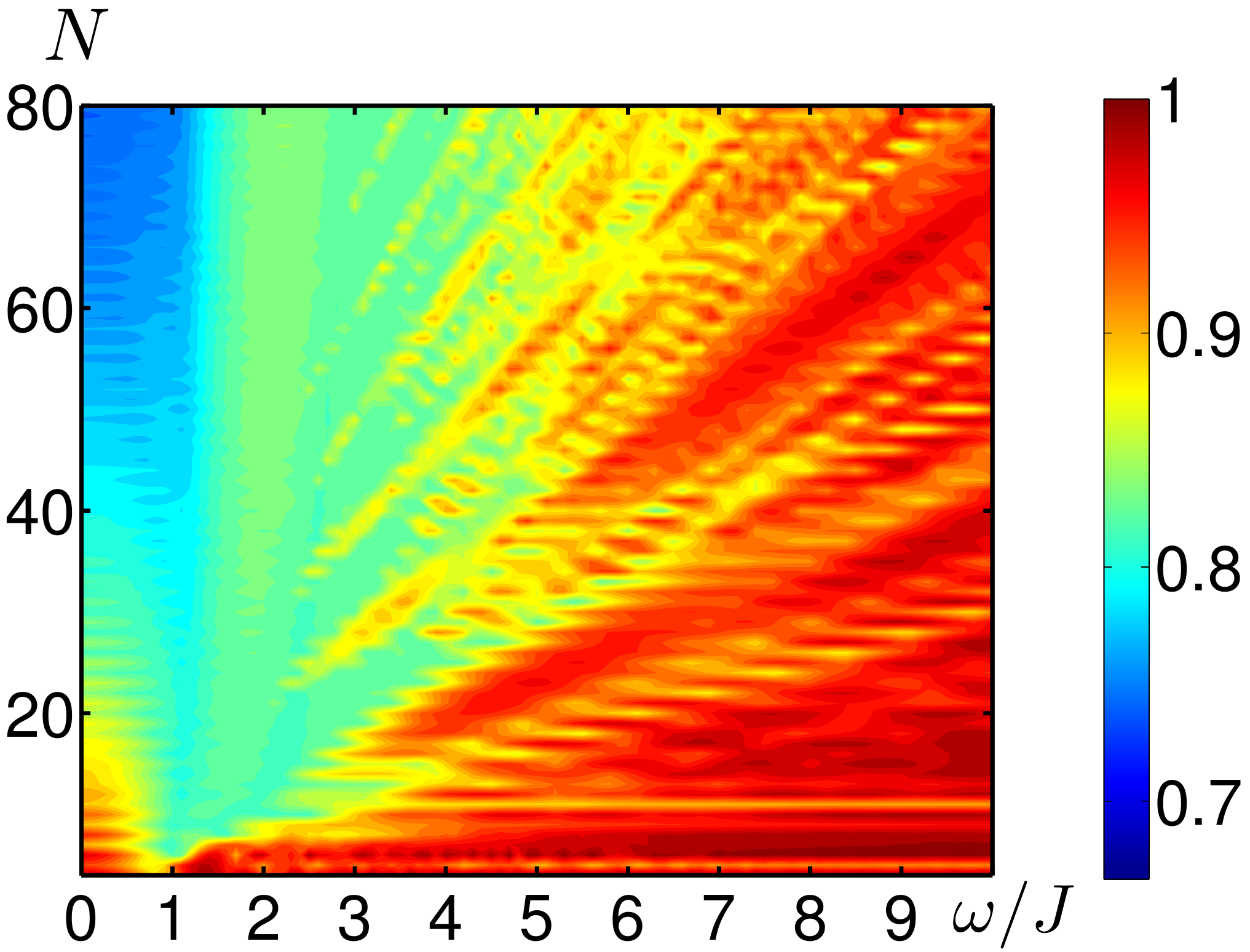}}
\caption{(color online): Maximum Fidelity in the time interval $J
t {\in}\left[0,4000\right]$ as function of $\omega$ (in units of
$J$) and $N$. Notice that for $\omega{=}0$ the Fidelity is larger
than $0.9$ only for short chains, while, as $\omega$ increases,
the Fidelity is significantly enhanced.} \label{FidelwN}
\end{figure}

In Fig.~\ref{FidelwN} we report the density plot of the maximum
Fidelity as a function of the number of sites and intensity
$\omega$ of the local fields to show that even modest values of
$\omega$ are sufficient for high-fidelity state transfer.

\begin{figure}[ht!]
    \center{{\bf (a)}\hskip4cm{\bf (b)}}
\center{\includegraphics[width=\linewidth]{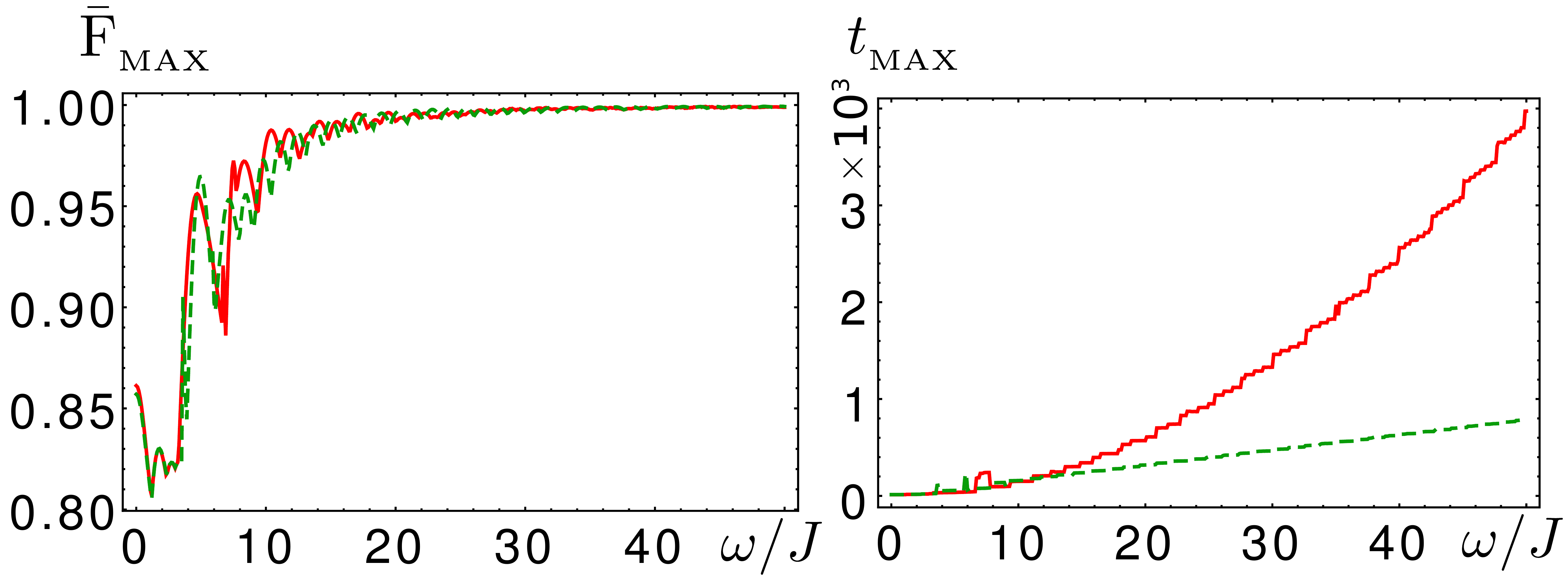}}
\caption{(color online): {\bf a}) Maximum Fidelity achievable in
the time interval $J t{\in}\left[0,4000\right]$. {\bf b}) Optimal
times at which the best transmission is attained. The plots refer
to chains of $N=22$ (red solid line) and $N=23$ sites (green
dashed line). For odd (even) $N$, $t_{\mathsf{MAX}}$ is linear
(quadratic) in $\omega$.} \label{Pari-dispari}
\end{figure}

By increasing $\omega$, the localization effect is enhanced and,
as a result, a better quantum-state transfer is obtained. This is
demonstrated in Fig.~\ref{Pari-dispari}, where the attainable
Fidelity tends towards $1$ both for even and for odd site numbers.
Nevertheless, as the eigenvalues of the bi-localized eigenvectors
become closer and closer to each other, by increasing $\omega$,
the transfer time increases. Since the transfer is based on
Rabi-like oscillations between the two eigenvectors with
IPR${\simeq}2$, the transfer time $t_{\mathsf{MAX}}$ can be
obtained from their eigenvalues:
$t_{\mathsf{MAX}}{=}\pi/(\lambda_2-\lambda_1)$, where
$\lambda_2{>}\lambda_1$. Furthermore, as shown by a
straightforward perturbation analysis, the eigenvalue difference
scales as $(N \omega)^{-1}$ for odd site numbers, while it behaves
as $\omega^{-2}$ for even ones, resulting in  shorter transfer
times for odd $N$~(see Fig.~\ref{Pari-dispari}{\bf b}). Notice
that the optimal transfer time does not directly depend on $N$ for
even site numbers, but $\omega$ needs to be increased~(almost
linearly) with increasing $N$ in order to have a Fidelity that
stays close to unity.
\subsection{Effective Hamiltonian description}
In this section, we compare our results with those obtained using
weak-end bonds~\cite{Wojcik2005}. To this end, we consider a
uniform magnetic field applied to a chain with sender and receiver
sites coupled more weakly to their neighboring spins than the
other nearest neighboring sites. Such a week bond is characterized
by an interaction strength $J^{\prime}$, being smaller than the
intra-chain exchange $J$. It turns out that a large Fidelity can
be obtained provided the ratio $J^{\prime}/J$ is suitably reduced
with increasing the chain's length. Moreover, with weak end-bonds,
a similar behavior of the transfer time is obtained, with an
even/odd asymmetry akin to the one discussed above.

The similarity is explained by observing that the magnetic field
barriers on the second and last-but-one spins give rise to
effective weak-end bonds, which, however, display some differences
with respect to the set-up of Ref.~\cite{Wojcik2005}. From a
perturbation analysis in terms of the small parameter $J/\omega
{\ll} 1$, we infer that the main effect of the local fields is to
modify the exchange interaction strengths between pairs of spins
near the sender and receiver sites. Indeed, the effective
Hamiltonian for the first three spins of the chain reads
\begin{equation}
H_{\mbox{eff}}{=}-
\left(\lambda_+
\ket{\psi_+}\bra{\bf 3}-\lambda_{-}\ket{\psi_-}\bra{\bf 3}+
\mbox{h.c.}\right).
\end{equation}
where, up to normalization factors, $\ket{\psi_+} {\propto}
\lambda_- \ket{\bf 1}{+}\ket{\bf 2}$,  $\ket{\psi_-} {\propto}
\lambda_+\ket{\bf 1}{+}\ket{\bf 2}$, and $\lambda_{\pm}{=}(\omega
{\pm} \sqrt{\omega^2+1})$. In the $\omega/J {\to} \infty$-limit, we
get $\ket{\psi_+} {\rightarrow} \ket{\bf 2}$ and $\ket{\psi_-}
{\rightarrow} \ket{\bf 1}$, so that the leading effect of the local
fields is the appearance of effective couplings $J_{13}$ and
$J_{23}$ between the corresponding spins. The latter are given by
\begin{equation*}
J_{13} \simeq
-\frac{1}{2 \omega} \qquad  J_{23} \simeq -\frac{1}{2}\left(1{-}\frac{1}{\omega^2}\right).
\end{equation*}
Summarizing, the effective hamiltonian of the first three spins of
the chain becomes $H_{\mbox{eff}}{=}J_{13} \left(\sigma^x_1
\sigma^x_3{+}\sigma^y_1 \sigma^y_3\right)+J_{23} \left(\sigma^x_2
\sigma^x_3{+}\sigma^y_2
\sigma^y_3\right){+}\lambda_-\sigma^z_1{+}\lambda_+\sigma^z_2$;
moreover, due to the presence of the large magnetic field on spin
2, its dynamics is frozen in the $\ket{0}$ state. Similar results
hold for the spins near the receiver.

Once the spins at sites $2$, $N-1$ are adiabatically eliminated,
we are effectively left with a chain of $N{-}2$ spins in a zero
magnetic field, uniformly coupled but for the end-bonds, where the
(effective) couplings between the spins $(1,3)$ and $(N-2,N)$ have
strength $J_{13}$.

A further perturbative analysis in the $J_{13}{\ll}1$ limit,
performed along the lines of Ref.~\cite{Wojcik2005}, allow us to
write an overall effective hamiltonian involving the spin-up
states at the sending and receiving sites only. More precisely,
this is strictly true only if $N$ is even; for a chain with an odd
number of sites, instead, the inclusion of an auxiliary state is
necessary, corresponding to the zero-energy eigenstate, whose
effects have been discussed in Section \ref{Sec.Model}.

As a result, for $N$ even and odd, respectively, the state
transfer is described by the following effective Hamiltonians:
\begin{align}\label{Heff}
H_{\mbox{eff}}^{even}{=}&
-\!\left(\frac{1}{4 \omega^2}\ket{\mathbf{1}}\!\bra{\mathbf{N}}+\mbox{h.c.}\right) \\
H_{\mbox{eff}}^{odd}{=}&
\frac{1}{2\omega}(1-\frac{4}{N-3})(\ket{\mathbf{1}}\!\bra{\mathbf{1}}+\ket{\mathbf{N}}\!\bra{\mathbf{N}}){+}\\
&\nonumber -\sqrt{\frac{2}{N-3}}\omega\left(\ket{\mathbf{1}}\!\bra{a}_{\frac{N{+}1}{2}}+\ket{\mathbf{N}}\!\bra{a}_{\frac{N{+}1}{2}}+\mbox{h.c.}\right)
\end{align}

\subsection{Robustness against Noise}
\label{SubSec.Noise} 
\begin{figure}[ht!]\centering
\includegraphics[width=\linewidth]{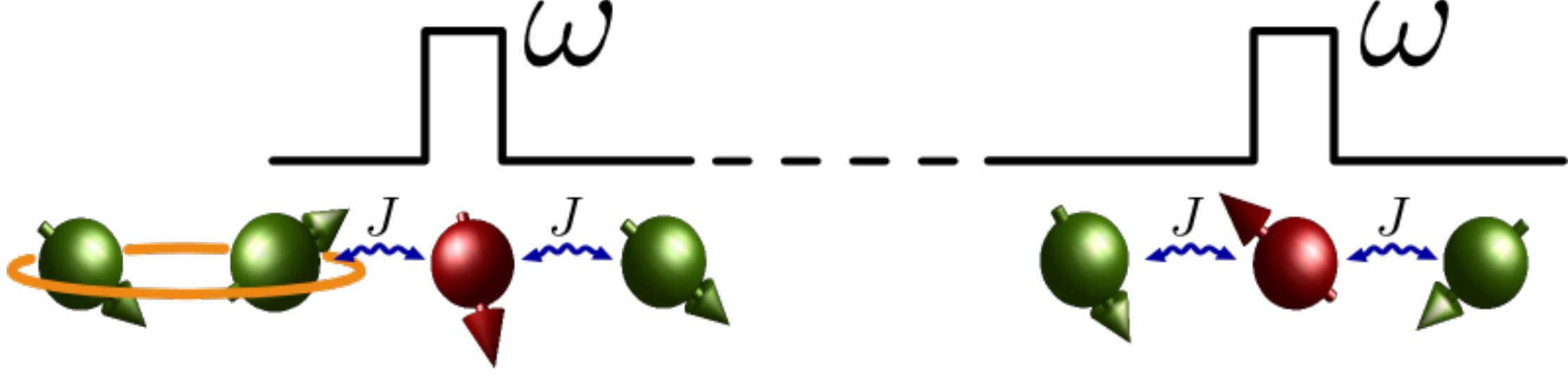}
\caption{(color online): Scheme of the set-up used for
entanglement transfer from the qubit pair $(0,1)$ to the pair
$(0,N)$, where the qubit $0$ is decoupled from the chain.}
\label{sketch-ent}
\end{figure}

In this subsection we investigate how a {\it
static} disorder in the magnetic fields acting on the qubits
$n{=}3,..,N{-}2$ affects the efficiency of information transfer,
and in particular, to be specific, of Entanglement transfer,
performed according to the scheme depicted in
Fig.~\ref{sketch-ent}. In this setting the Entanglement, initially
contained in the state $\ket{\Psi_+}{=}
\frac{1}{\sqrt{2}}(\ket{01}{+}\ket{10})$ of the qubit pair
$(0,1)$, is transferred to the pair $(0,N)$ and is quantified by
the Concurrence as given by Eq.~\eqref{ConC}.

\begin{figure}[ht!]\label{disorder}
\includegraphics[width=\linewidth]{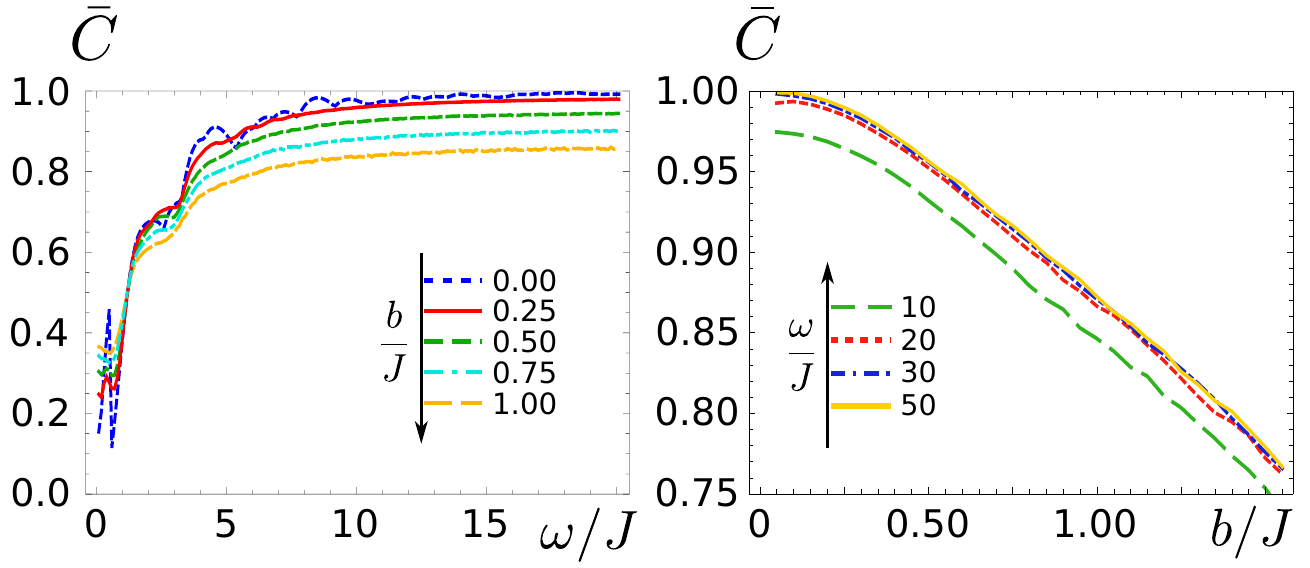}
\caption{(color online): (left panel) Averaged Concurrence vs
auxiliary field strength $\omega$ for different values of the
disorder parameter $b$. (rigth panel) Concurrence vs disorder
strength $b$ for different values of $\omega$. The curves for
$\omega{>}20$ are almost indistinguishable and no relevant change
in the effects of the disorder is observed. In both panels, the
length of the chain is $N{=}10$ and averages are performed over
$10^{5}$ realizations of disorder. All energy values are expressed
in units of $J$. }
\end{figure}

The kind of disorder we consider is given by the presence of
random local magnetic fields between the barrier qubits. In other
words, we are assuming that local random magnetic fields,
uniformly distributed in an interval $-b < K_n < b$, with $b$
denoting the disorder strength, act on the spins residing on sites
$n = 3, \ldots , N{-}2$. This choice is justified by the fact that
the hamiltonian parameter of the qubits $(1,N)$ are generally
considered to be more precisely controllable in order to perform
efficiently the state encoding and read-out procedure and,
therefore, they will be practically unaffected by the disorder.
Furthermore, we allow the same degree of control for the
neighboring spins $2,N-1$, whose local fields are assumed to be
precisely fixed. In Figs.~\ref{disorder}, we see that the
attainable Concurrence~\eqref{ConC}, averaged over $10^{5}$
samples of disorder, remains quite high provided that
$b{\ll}\omega$. Indeed, the bi-localized nature of the relevant
eigenstates is not significantly perturbed. On the contrary, this
is not anymore the case for values of $b$ comparable to- or
greater than- $\omega$. Similar results are obtained for the
Fidelity of the QST.

\begin{figure}
\includegraphics[width=\linewidth]{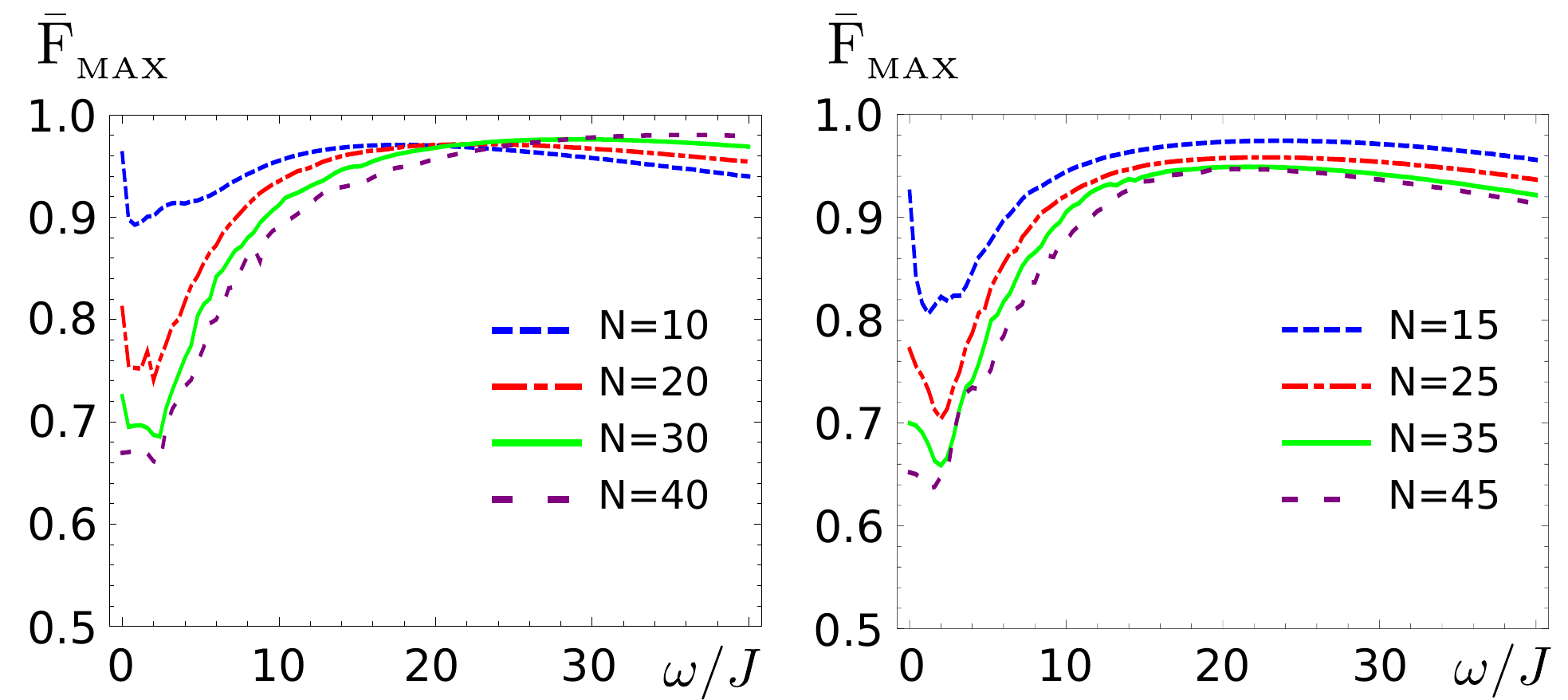}
\caption{(color online): Averaged Fidelity vs auxiliary field
strength $\omega$ for different lengths of the chain. In this case
a residual magnetic field is supposed to act on sites $3,N-2$ and
$4,N-3$, with random values uniformly distributed between $0$ and
$\omega/10$ for the sites near the barriers, and between $0$ and
~$\omega/40$ the the next to nearest sites, respectively. The
plots show averages performed over $10^{5}$ realizations. }
\label{nuovodis}
\end{figure}

Depending on the specific physical implementation of the model,
other sources of errors (and, specifically, of static disorder)
can be identified. In particular, we would like to mention that
the robustness of different transfer schemes against bond
disorder, (that is, static disorder in the spin-coupling
strengths) has been investigated in Ref.~\cite{ZwickASO2011}. It
turns out that the localization properties of the eigenstates play
an important role for efficient state transfer in presence of
non-uniform bonds, and that a mechanism based on localized states,
like the one we are describing, is more resilient then a ballistic
transport-based one.
On the other hand, since we consider high magnetic field applied
locally to sites $2$ and $N-1$, a leakage effect is certainly
possible, affecting the neighboring sites. To check the robustness
of our transfer scheme against this lack of control, we can
consider random magnetic fields, with amplitude decaying with the
distance, to affect the dynamics of spins $3,4, N-3,N-2$ (on the
other hand, as discussed above, we assume a very high degree of
control on the sending and receiving sites, and on the barrier
fields). The results of such an analysis are reported in Fig.
\ref{nuovodis}, where the transmission fidelity averaged over
$10^5$ realization of these static random fields is displayed. For
very small values of the local fields $\omega$, the quality of the
transfer is strongly reduced by the presence of this kind of
disorder, while its effect is shown to substantially decrease for
larger values of barrier fields, despite the residual static
random fields are bounded always by the same fractions of
$\omega$.

The plots suggest that, both for chains with odd and even $N$, an
optimal value of the local barrier fields exists in the case in
which a given fraction of it is assumed to leak to the neighboring
sites. If such an optimal value of $\omega$ is selected (which
scales almost linearly with the size $N$), the average fidelity is
kept very close to unity.

\subsection{Transport of an entire e-bit}
\label{SubSec.Twoqubit} 
\begin{figure}[ht!]
    \centering
   \includegraphics[width=\linewidth]{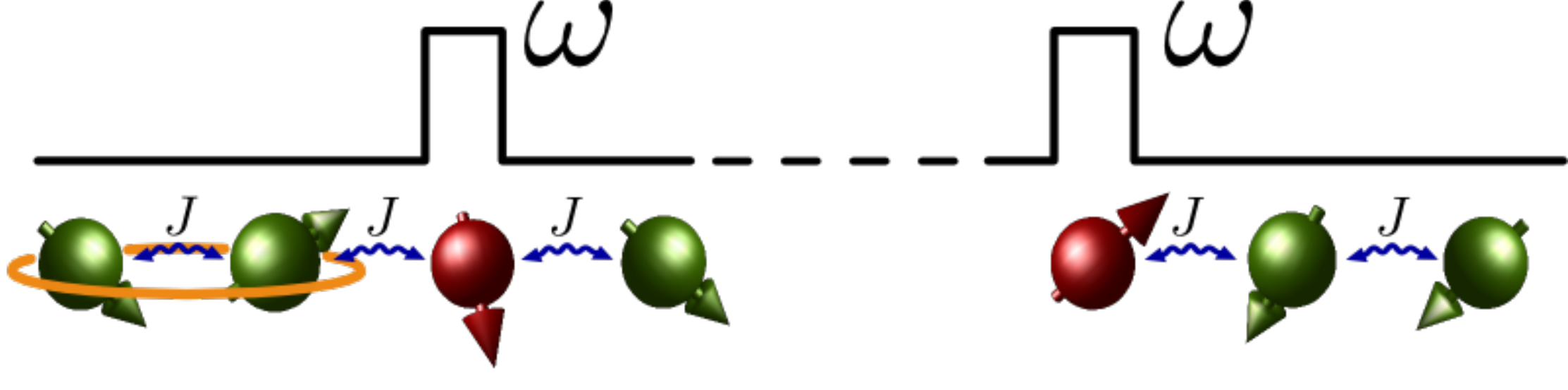}
\caption{(color online): Sketch of the configuration for the
transfer of an e-bit.}\label{2tr-chain}
\end{figure}

\begin{figure}[ht!]
    \centering
  \includegraphics[width=\linewidth]{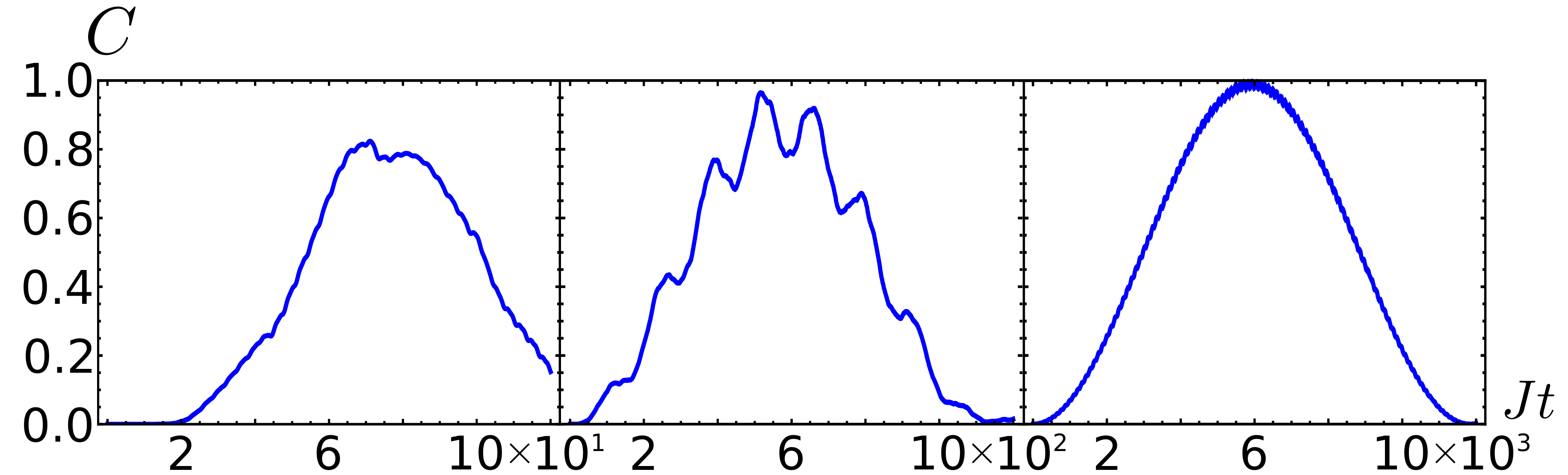}
\caption{(color online): Concurrence transferred from the
maximally initial entangled qubits pair $(1,2)$ to the pair
$(N{-}1,N)$ for a chain of $N{=}33$ and $\omega{=}5,15,45$ (from
left to right). Higher values of the local  field, beside
increasing the amount of transferred Entanglement, regularize the
dynamics.}
  \label{twoqubit}
  \end{figure}

We have shown above that a qubit encoded
on the first spin of the chain is almost perfectly transferred to
the other end, thanks to the application of local magnetic fields
to the adjacent spins to the sender and the receiver sites. In
this subsection we extend this idea to the transfer of an
entangled pair. Considering the setup depicted in
Fig.~\ref{2tr-chain}, we aim at transferring the Entanglement
shared by qubits $1$ and $2$ to qubits $N{-}1$ and $N$ by use of
auxiliary magnetic fields applied to sites $3$ and $N{-}3$. We,
thus, allow  $K_n{=}\omega\left(\delta_{n,3}+\delta_{n,N-2}
\right)$ in Eq.~(\ref{H-general-local}).
Then, we start from the fully polarized state
$\ket{{\mathbf{0}}}$, and initialize the first two spins in a
state belonging to the single excitation subspace so that the
initial state of the whole chain reads:
\begin{equation}
|\Psi(0)\rangle = \alpha |\textbf{1}\rangle +
\beta|\textbf{2}\rangle\;\;\;\;\;\;(\left|\alpha\right|^2+\left|\beta\right|^2)=1,
\end{equation}
whose evolution is given by
\begin{equation}
\ket{\Psi(t)}{=}\sum_{j=1}^{N}p_{j}|\mathbf{j}\rangle ,\quad
p_{j}=\alpha \bra{\textbf{j}}e^{-iHt} \ket{\textbf{1}}{+}\beta
\bra{\textbf{j}}e^{-iHt} \ket{\textbf{2}}.
\end{equation}
Finally, we obtain the state of the qubits $N{-}1$ and $N$ by
performing the partial trace over the first $N{-}2$ spins.
Considering an initially entangled $(1,2)$~pair (that is,
$\alpha,\beta{\neq}0$), the amount of entanglement transferred to
the pair $(N-1,N)$ and measured by the Concurrence is given by
$C_{N-1,N}{=}2|p_{N-1}p_N|$. As shown in Fig.~\ref{twoqubit} where
an initial maximally entangled state has been taken, i.e.,
$\alpha{=}\beta{=}\frac{1}{\sqrt{2}}$, also the Entanglement may
be efficiently transferred in the presence of the auxiliary
magnetic fields.

\section{Time-dependent Quantum State Transfer Protocol}
\label{Sec.Time} In this Section we investigate a QST protocol in
which we allow for time-control of the magnetic fields acting on
the barrier qubits. The aim of this control is to provide a
precise timing for the beginning and end of the sending stage, as
given by the switching of the local fields. At the same time, the
control relaxes the need of a fast~(in fact, instantaneous)
extraction of the received information at the site $N$ once the
transmission is performed. The idea is to encode the quantum state
on the sender site and leave it there for a future transmission by
means of a strong magnetic field on its neighbor barrier qubits.
In this first step the information stays localized on the sender
as no tunnelling of the spin excitation is possible due to the
energy mismatch with other sites. The sending stage is then
realized by switching on the magnetic field of the other barrier,
at the $(N-1)$-th site. During this second step, the Rabi
oscillation described in the previous Sections takes place.
Finally, in the third stage, only the barrier on the last-but-one
spin is left on, in order to trap the received quantum state,
while the local field near the sender site is switched off.

To implement this proposal, we exploit the time-dependent Hamiltonian
\begin{equation}\label{H-general-time}
\mathcal{H}(t)=-\dfrac{1}{2}\sum_{n=1}^{N-1}
(\sigma_{n}^x\sigma_{n+1}^x{+}\sigma_{n}^y\sigma_{n+1}^y)
{-}\sum_{i=2,N{-1}}\omega_i(t) \sigma_{i}^z,
\end{equation}
where
\begin{equation}\begin{array} {ccc}
\omega_2(t){=}
\begin{cases}
 K_1 & \\ 
 K_2 & \\ 
 0 & \\ 
\end{cases}
\end{array}
\quad
\begin{array} {ccc}
\omega_{N-1}(t) {=}
\begin{cases}
 0& t_0{\leq}t{<}t_1 \\
 K_2& t_1{\leq} t{\leq}t_2 \\ 
 K_1& t{>}t_2 
\end{cases}
\end{array}
\label{omegasteps}
\end{equation}
Here, $t_2{=}t_1{+}\Delta t$, with $\Delta t$ being an optimal
transfer time interval, that we define below.
\begin{figure}[ht!]
     \includegraphics[width=6cm]{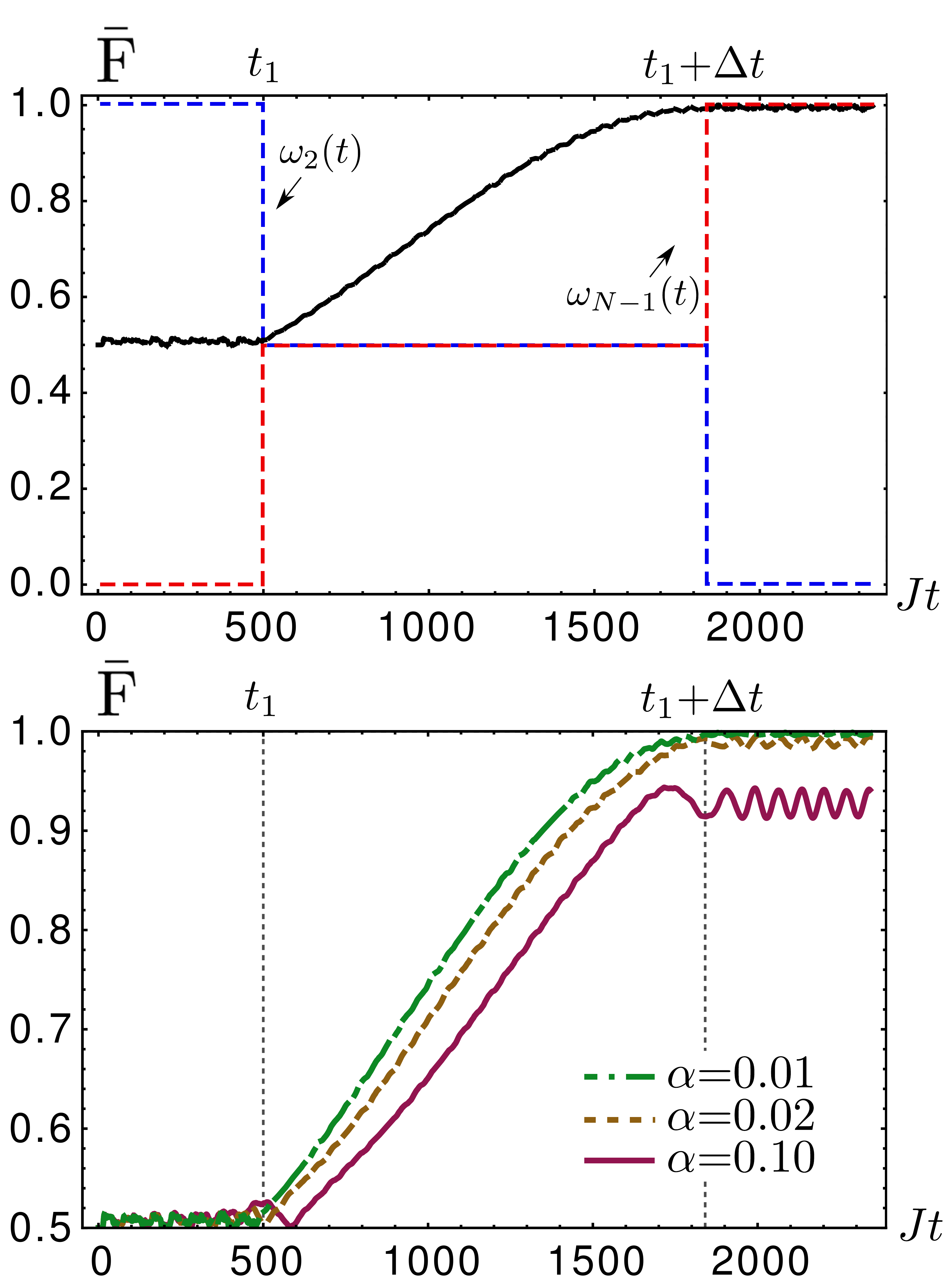}
\caption{(color online): (upper panel) Average Fidelity for a
chain of $N{=}30$ sites with $K_1{=}60$ and $K_2{=}30$, where the
three steps of the QST-protocol are clearly visible from the time
behavior of the two fields, which are switched according to the
recipe of Eq.~(\ref{omegasteps}). (lower panel) Same as in the
upper plot, but with finite switching times for the fields. In
this case, $\omega_2(t)$ and $\omega_{N-1}(t)$ are smother
versions of the step functions of Eq. (\ref{omegasteps}), with
exponential corrections: $\omega_2(t) = K_2 /( \exp \{\alpha
(t-t_2) \} + 1) + (K_1-K_2) /( \exp \{\alpha (t-t_1) \} + 1)$, and
a similar behavior for $\omega_{N-1}(t)$. The three curves
correspond to three values of $\alpha$ and it turns out that the
achievable ${\overline{F}}$ decreases with the steps becoming
smoother and smoother (that is, with increasing $\alpha$).}
\label{dyn12}
\end{figure}
With this time-dependent field configuration, the spin at the
first site is ``frozen'' until $t{<}t_1$ as the state
$\ket{\mathbf{1}}$ is an approximate eigenstate of
$\mathcal{H}(t{<}t_1)$; then, after the resonant tunnelling to the
receiving site $t_1{\leq} t{\leq} t_1{+}\Delta t$, for $t{>}t_2$,
the information is definitely stored in the $N$-th spin, as
$\ket{\mathbf{N}}$ is an approximate eigenstate of
$\mathcal{H}(t{>}t_2)$~[see the upper panel in Fig.\ref{dyn12}].

Since the transition amplitude is given by $\bra{N}U(t)\ket{1}$,
in order to obtain the Fidelity, one needs to solve the time
dependent Schr\"{o}dinger equation for the state
$U(t)\ket{1}{=}\sum_{k{=}1}^N\beta_k(t)\ket{k}$ that reduces to an
$N{\times} N$ system of differential equations:
\begin{equation}\begin{cases}
i\dfrac{d \beta_1(t)}{dt}{=}-\beta_2(t)\\
i\dfrac{d \beta_2(t)}{dt}{=}-\beta_1(t)-\omega_2(t)\beta_2(t)-\beta_3(t)\\
...\\
i\dfrac{d \beta_j(t)}{dt}{=}-\beta_{j-1}(t)-\beta_{j+1}(t)\;\;\;\;(j=3,...,N-2)\\
...\\
i\dfrac{d \beta_{N-1}(t)}{dt}{=}-\beta_{N-2}(t)-\omega_{N-1}(t)\beta_{N-1}(t)-\beta_N(t)\\
i\dfrac{d \beta_{N}(t)}{dt}{=}-\beta_{N-1}(t).\\
\end{cases}\end{equation}
The solution, in each time interval where $\omega_2(t)$ and $\omega_{N-1}(t)$ are constant, is
\begin{equation}
\beta_s(t_{i+1}){=}(-1)^{s+1} \sum_{\lambda_j }\sum _{k=1}^N
\frac{Q_{k,s}(\lambda )} {\frac{dP(\lambda)}
{d\lambda}}\vert_{\lambda_j} e^{-i\lambda_j t}\beta_k(t_{i}).
\end{equation}
where $\lambda_j$ are the eigenvalues of $H(t_i)$, while
$P(\lambda)$ and $Q_{k,j}(\lambda)$ are, respectively, the
determinant and the minors of the matrix $(H(t_{i+1})-\lambda
\unit) $. In order to perform the state transfer along the chain,
the optimal time $\Delta t$ is proportional, again, to the inverse
of the eigenvalues difference of the intermediate stage
Hamiltonian, and reads
\begin{equation*}
\Delta t{=}
\begin{cases}
\dfrac{\pi}{2}K_2^2\qquad \qquad \qquad N\;\;even\\ \\
\dfrac{\pi}{4}(N-3)K_2\qquad \;\;N\;\;odd.\end{cases}
\end{equation*}
The procedure works quite well for even $N$, as illustrated in
Fig. \ref{dyn12}, where the detrimental effect of finite switching
times for the field is also explored. Unlike adiabatic transfer
schemes~\cite{Chenetal12}, here a fast switching of the magnetic
fields is desirable because the overlap of the initial (final)
state $\ket{\mathbf{1}}$ ($\ket{\mathbf{N}}$) with the
bi-localized states is maximized by a step function, whereas a
smoother switching function would introduce into the dynamics
destructively interfering states that do not possess the required
localization properties. This is illustrated in the lower panel of
Fig.\ref{dyn12}, where the average Fidelity is plotted for
different switching rates.

For odd chains, on the other hand, the presence of the zero-energy
eigenstate taking part in the dynamics, makes the state transfer
more involved, because the trapping stages both at the beginning
and end of the protocol are not so efficient.
\section{Conclusions}
\label{Sec.Conc} Spin chain models describe a great variety of
different physical systems, ranging from trapped ions interacting
with lasers~\cite{PorrasC04}, via flux qubits \cite{vseivotto} and
arrays of coupled cavity, to ultra-cold atoms in 1D optical
lattices~\cite{GiampaoloI,treutreq}, including coupled quantum
dots~\cite{PetrosyanL06}, nitrogen vacancy centers in diamond
\cite{yao}, or magnetic molecules~\cite{Troiani09}. All these
possible implementations have their own strengths and weaknesses
and allow for different possible kinds of controls on the single
units. It is therefore of  interest to put forward QST protocols
that may fit better to a specific experimental realization of the
quantum channel.

In many of the above mentioned implementations, only a restricted
access is possible to the Hamiltonian parameters. It is therefore
desirable, to study efficient and reliable transmission protocols
that require only a limited amount of controls. In this paper we
have shown that a high-quality quantum state transfer can be
achieved in a $XX$-spin chain by means of strong local magnetic
fields applied on the second and last-but-one spins, that cause
appearance of two specific eigenstates, bi-localized on the sender
and receiver sites located at the edges of the chain. Unlike other
QST protocols, this implies that no engineering of the Hamiltonian
parameters is required. A much more limited control is needed only
on some local properties of the spins close to the sender and the
receiver sites. By increasing the magnetic fields $\omega$, the
transfer Fidelity has been shown to approach unity, with a
transfer time scaling as $\omega^{-1}$ and $\omega^{-2}$ for
chains with odd and even numbers of sites, respectively;
furthermore, a good resilience to the presence of static disorder
in the local Hamiltonian parameters of the channel has been
demonstrated. The model works also for the transfer of a two-qubit
state, and more general $n$-qubit state transmission can be easily
envisaged using similar schemes. Furthermore, this set-up allows
for an efficient time-dependent protocol, based on fast switching
of the magnetic fields, which has the benefit of avoiding the need
for a fast and well synchronized state retrieval. The latter is a
common requirement for many existing QST proposals. Indeed, with
our set-up, the transferred state can be trapped, with a high
Fidelity of storage, at the end of the transmission protocol, thus
allowing for a much easier extraction of the information.

\acknowledgments
TJGA is supported by the
European Commission, the European Social Fund and the Region
Calabria through the program POR Calabria FSE 2007-2013-Asse IV
Capitale Umano-Obiettivo Operativo M2.

\end{document}